\documentclass[11pt,onecolumn,aps,superscriptaddress,
               amsmath,amssymb,nofootinbib]{revtex4}      
\usepackage{graphicx}

\usepackage{graphicx,epsfig}

\usepackage{dcolumn}
\usepackage{bm}

\begin{document}

\begin{flushright}
\baselineskip=12pt \normalsize
{ACT-12-09},
{MIFP-09-47}\\
\smallskip
\end{flushright}

\title{A Note on Modulus-dominated SUSY-breaking}

\author{James A. Maxin}
\affiliation{George P. and Cynthia W. Mitchell Institute for
Fundamental Physics, Texas A\&M
University,\\ College Station, TX 77843, USA}
\author{Van E. Mayes}
\affiliation{Physics Department, Arizona State University,\\ Tempe, AZ 85287-4111, USA}
\author{D.V. Nanopoulos}
\affiliation{George P. and Cynthia W. Mitchell Institute for
Fundamental Physics, Texas A\&M
University,\\ College Station, TX 77843, USA}
\affiliation{Astroparticle Physics Group, Houston
Advanced Research Center (HARC),
Mitchell Campus,
Woodlands, TX~77381, USA; \\
Academy of Athens,
Division of Natural Sciences, 28~Panepistimiou Avenue, Athens 10679,
Greece}

\begin{abstract}
\begin{center}
{\bf ABSTRACT}
\end{center}
In models where supersymmetry-breaking is dominated by the K\a"ahler moduli and/or the universal dilaton, the B-parameter at the unification scale should be consistent with the value of tan$\beta$ at the electroweak scale determined by minimization of the Higgs potential triggering REWSB.  We study such models employing a self-consistent determination of the B-parameter. In particular, we study the viability of  a generic model, as well as M-theory and Type IIB flux compactifications with modulus-dominated supersymmetric soft-terms from the GUT scale, $M_{GUT}=2\times10^{16}$~GeV. 
\end{abstract}

\maketitle

\newpage
\section{Introduction}

If TeV-scale supersymmetry is discovered at LHC, it will open a window in which to explore physics at higher-energy scales.
In particular, the measurement of superpartner masses can provide a test of different proposed mechanisms for breaking supersymmetry.
Moreover, it may allow us to probe the underlying theory which provides the UV completion of known low-energy physics.  In particular, in various string theory compactifications where the effective low-energy $\mathcal{N}=1$ supergravity
approximation holds true, it is possible to 
generate superpartner spectra which may be compared to whatever may be observed at LHC.  

The most studied model of supersymmetry breaking is minimal supergravity (mSUGRA), which arises
from adopting the simplest ansatz for the K\a"ahler metric, treating all chiral superfields
symmetrically.  In this framework,~~$\mathcal{N}=1$ supergravity is broken in a hidden sector
which is communicated to the observable sector through gravitational interactions.  Such models
are characterized by the following parameters: a universal scalar mass $m_0$, a universal gaugino mass
$m_{1/2}$, the Higgsino mixing $\mu$-parameter, the Higgs bilinear $B$-parameter, a universal trilinear coupling $A_0$, and tan~$\beta$.  One then determines the $B$ and $|\mu|$ parameters by the minimization of the Higgs potential triggering REWSB, with the sign of $\mu$ remaining undetermined.  Thus, we are left with only four parameters. 

Although, mSUGRA is one of the most generic frameworks that can be adopted, many string compactifications
typically yield expressions for the soft terms which are even more constrained, in particular, when supersymmetry breaking is dominated by the K\a"ahler moduli and/or dilaton. As is well-known, the K\a"ahler moduli of Type I, IIB
orientifold, 
and heterotic string compactifications have a classical no-scale structure~\cite{DVNS,JLDN, VKJL, ABIM}, which guarantees that the vacuum energy vanishes at tree-level.  The no-scale structure corresponds to having non-vanishing expectation values for the auxiliary fields of the K\a"ahler moduli. 
The generic appearance of the 
no-scale structure across many string compactifications combined with the highly-constrained and thus predictive framework strongly motivates the consideration of modulus-dominated supersymmetry breaking, although there are some string models for which the soft-terms are not as constrained (see ~\cite{Chen:2007px, Chen:2007zu, Maxin:2009qq, Maxin:2009ez} for a model of this kind).

For modulus-dominated supersymmetry breaking, we generically have $m_{0} = m_{0}(m_{1/2})$ and $A = A(m_{1/2})$. This reduces the number of free parameters compared to mSUGRA down to two, $m_{1/2}$ and tan$\beta$.  In fact, adopting a 
strict no-scale framework, one can also fix the $B$-parameter as $B=B(m_{1/2})$, and thus we
are led to a {\it one-parameter} model where all of the soft terms may be fixed in terms of 
$m_{1/2}$.  However, for this framework to be consistent, the value of tan$\beta$ at the electroweak 
scale should be consistent with $B$ at the string scale.    

In a previous paper, we studied a generic one-parameter model and found its viable parameter space~\cite{Maxin:2008kp}.  However, in this work we did
not require that tan$\beta$ obtained at the electroweak scale be consistent with the value of $B=B(m_{1/2})$
defined at the GUT scale. For the present work, we impose this constraint for a generic one-parameter model and find that there is no viable supersymmetry parameter space, assuming the standard RGE running between the electroweak scale and the GUT scale.  Furthermore, we find the same result for M-theory and Type IIB flux compactifications. In addition, we consider different modular weights for some of the chiral fields, again with negative results.  We conclude that modulus-dominated supersymmetry
breaking is not viable, in the case of a standard RGE running of the soft terms starting from the GUT scale.   

\section{Modulus-dominated SUSY-breaking}

For certain classes of string compactifications, the soft-terms are of the form
$m_{0} = m_{0}(m_{1/2})$ and $A = A(m_{1/2})$ if supersymmetry is dominated by the K\a"ahler moduli and/or the universal dilaton.  In particular,
much work has been done in the past to study two generic cases inspired by no-scale supergravity in the framework of the free-fermionic class of heterotic string compactifications.   
The first
of these two cases is referred to as the special {\it dilaton} scenario,
\begin{equation}
\label{eqn:dilaton}
m_{0} = \frac{1}{\sqrt{3}}m_{1/2}, \ \ \ \ \ A = -m_{1/2}, \ \ \ \ \ B = \frac{2}{\sqrt{3}}m_{1/2}.
\end{equation}
while the second is referred to as the strict {\it moduli} scenario,
\begin{equation}
\label{eqn:moduli}
m_{0} = 0, \ \ \ \ \ A = 0, \ \ \ \ \ B = 0.
\end{equation}
In previous work, it was found that there is no viable parameter space for the strict moduli scenario which satisfies experimental constraints.
However, in the case of the special dilaton scenario there is a small allowed parameter space.  

Moreover, the soft-terms for many string compactifications will also be of similar form.  
In particular, the soft terms for heterotic M-theory compactifications take
the form~\cite{TJLI}
\begin{eqnarray}
&m_{1/2} = \frac{x}{1+x}m_{3/2}, \\ \nonumber
&m_{0} = \frac{x}{3+x}m_{3/2}, \\ \nonumber
&A = - \frac{3x}{3+x}m_{3/2}, 
\end{eqnarray}
\noindent while for dilaton dominated supersymmetry breaking they take the form
\begin{eqnarray}
&m_{1/2} = \frac{\sqrt{3}m_{3/2}}{1+x}, \\ \nonumber
&m_{0}^{2} = m_{3/2}^{2} - \frac{3m_{3/2}^{2}}{(3+x)^{2}}x(6+x), \\ \nonumber
&A = - \frac{\sqrt{3}m_{3/2}}{3+x}(3-2x). 
\end{eqnarray}
\noindent These expressions reduce to the above moduli and dilaton scenarios respectively in the limit $x \rightarrow 0$,
where
\begin{equation}
x \propto \frac{(T+\overline{T})}{S+\overline{S}}
\end{equation}

In addition, the so-called large-volume models have been studied extensively~\cite{BBCQ}~\cite{JCQS} in recent years and the generic soft terms for this framework have been calculated in~\cite{CAQS}. These models involve Type IIB compactifications where the moduli are stabilized by fluxes and quantum corrections to the K\a"ahler potential generate an exponentially large volume. This exponentially large volume may lower the string scale to an intermediate level which can be in the range $m_{s} \sim 10^{3-15}$ GeV. In such models, the
soft terms can take the form
\begin{eqnarray}
 \centering
	&m_0 = \frac{1}{\sqrt{3}}M, \nonumber \\
	&A_0 = -M, \nonumber \\
	&B = - \frac{4}{3}M,
\end{eqnarray}
where $M$ is a universal gaugino mass.

As can be seen for these different string compactifications, the soft terms can generically be of the form
\begin{eqnarray}
m_0 = c_1 m_{1/2}, \nonumber \\
A_0 = c_2 m_{1/2}, \nonumber \\
B =   c_3 m_{1/2},
\end{eqnarray}
where $c_1$, $c_2$, and $c_3$ are constants. In addition, we will
take the string scale to be $M_{GUT}=2\times 10^{16}$~GeV.  However, we should note that the string scale at which the soft-terms are defined
could be different from the conventional GUT scale.  In particular, we can see for the case of the M-theory compactifications, the unification scale
can be higher than the GUT scale, while for the large-volume Type IIB flux compactifications, the string scale could be substantially lower.

\section{Imposing the B Constraint}
As stated in the introduction,
the value of the $\mu$ parameter and tan$\beta$ are determined at the electroweak scale by imposing the conditions
\begin{equation}
\mu^2 = \frac{-m^2_{H_u}\mbox{tan}^2\beta+m^2_{H_d}}{\mbox{tan}^\beta -1} - \frac{1}{2}M^2_Z,
\label{higgsmin1}
\end{equation}
and 
\begin{equation}
\mu B = \frac{1}{2}\mbox{sin}2\beta(m^2_{H_d}+m^2_{H_u}+2\mu^2),
\label{higgsmin2}
\end{equation}
which follow from the minimization of the Higgs potential triggering REWSB.  From these equations,
one can calculate the value of the B-parameter at the electroweak scale.  In order for this to be a true one-parameter
model, $B$ at the electroweak scale should be consistent with the ansatz $B=B(m_{1/2})$ at the GUT scale. 

The usual procedure to find the viable parameter space is to calculate the sparticle masses using the parameters $m_{0}$, $m_{1/2}$, $A_{0}$, $sgn(\mu)$, and tan$\beta$, and plot $m_{0}~ vs.~ m_{1/2}$ for a specific tan$\beta$, and further scan the entire tan$\beta$ space for solutions that satisfy the current experimental constraints and corresponding relic neutralino density. In particular, such an analysis was performed for a generic one-parameter model in~\cite{Maxin:2008kp}.  However, the consistency constraint between the B-parameter at the electroweak scale and the GUT scale has not been imposed in this analysis. For the present work, we perform a scan of the parameter space, including tan$\beta$, and filter the results through the latest experimental constraints and dark matter density, and in addition, compare the allowed parameter space with the value of the B-parameter at $M_{High}$. For the present work, we will identify $M_{High}$ with $M_{GUT}$.  This determines whether the allowed parameter space calculated from tan$\beta$ can also satisfy the constraint on the B-parameter at the unification scale (see~\cite{Aparicio:2008wh} for a similar study in the case of F-theory compactifications).

First, we generate sets of soft supersymmetry breaking terms at the unification scale for the models we consider, then the soft terms are input into {\tt MicrOMEGAs 2.0.7}~\cite{Belanger:2006is} using {\tt SuSpect 2.34}~\cite{Djouadi:2002ze} as a front end to evolve the soft terms down to the electroweak scale via the Renormalization Group Equations (RGEs) and then to calculate the corresponding relic neutralino density. We take the top quark mass to be $m_t = 173.1$~GeV~\cite{:2009ec} and leave tan~$\beta$ as a free parameter, while $\mu$ is determined by the requirement of REWSB. However, we do take $\mu > 0$ as suggested by the results of $g_{\mu}-2$ for the muon. The resulting superpartner spectra are filtered according to the following criteria: 
 
\begin{enumerate}

\item The 5-year WMAP data combined with measurements of Type Ia supernovae and baryon acoustic oscillations in the galaxy distribution for the cold dark matter density~\cite{Hinshaw:2008kr},  0.1109 $\leq \Omega_{\chi^o} h^{2} \leq$ 0.1177, where a neutralino LSP is the dominant component of the relic density. In addition, we look at the SSC model~\cite{Antoniadis:1988aa}, in which a dilution factor of $\cal{O}$(10) is allowed~\cite{Lahanas:2006hf}, where $\Omega_{\chi^o} h^{2} \lesssim$ 1.1. For a discussion of the SSC model within the context of mSUGRA, see~\cite{Dutta:2008ge}. We also investigate another case where a neutralino LSP makes up a subdominant component, allowing for the possibility that dark matter could be composed of matter such as axions, cryptons, or other particles. We employ this possibility by removing the lower bound.

\item The experimental limits on the Flavor Changing Neutral Current (FCNC) process, $b \rightarrow s\gamma$. The results from the Heavy Flavor Averaging Group (HFAG)~\cite{Barberio:2007cr}, in addition to the BABAR, Belle, and CLEO results, are: $Br(b \rightarrow s\gamma) = (355 \pm 24^{+9}_{-10} \pm 3) \times 10^{-6}$. There is also a more recent estimate~\cite{Misiak:2006zs} of $Br(b \rightarrow s\gamma) = (3.15 \pm 0.23) \times 10^{-4}$. For our analysis, we use the limits $2.86 \times 10^{-4} \leq Br(b \rightarrow s\gamma) \leq 4.18 \times 10^{-4}$, where experimental and
theoretical errors are added in quadrature.

\item The anomalous magnetic moment of the muon, $g_{\mu} - 2$. For this analysis we use the 2$\sigma$ level boundaries, $11 \times 10^{-10} < a_{\mu} < 44 \times 10^{-10}$~\cite{Bennett:2004pv}.

\item The process $B_{s}^{0} \rightarrow \mu^+ \mu^-$ where the decay has a $\mbox{tan}^6\beta$ dependence. We take the upper bound to be $Br(B_{s}^{0} \rightarrow \mu^{+}\mu^{-}) < 5.8 \times 10^{-8}$~\cite{:2007kv}.

\item The LEP limit on the lightest CP-even Higgs boson mass, $m_{h} \geq 114$ GeV~\cite{Barate:2003sz}.

\end{enumerate}

To determine the B-parameter at $M_{High}=M_{GUT}$, B is determined at $m_{Z}$ from Eqns. (\ref{higgsmin1}) and (\ref{higgsmin2}). Then it is run up to the unification scale to compute the boundary condition for B. A sufficient number of iterations between $m_{Z}$ and $m_{GUT}$ are calculated until stable results are achieved. The value for B at the GUT scale we use is at the last iteration before the results become stable. To accomplish this, we modify the {\tt SuSpect} code to output the B-parameter value from the RGE loop during this final iteration. We capture the B-parameter through this method for all sets of the soft-supersymmetry breaking terms that we calculated the experimentally allowed parameter space. Once B is computed for all points, we compare this value of B to the theoretical prediction for B at the unification scale for each model we consider in a plot of the ratio of B to the gaugino mass versus tan$\beta$. The only points that satisfy the B constraint are those points on the B-parameter curves that intersect with the horizontal lines representing the theoretical prediction. Additionally, it is also necessary for these points of intersection between the B curves and predictions to lie within the range of points within the experimentally allowed parameter space. These points just described will satisfy not just the aforementioned five experimental constraints, but also the constraint on the B-parameter at the unification scale. However, as we will show here, it is very difficult to satisfy all these constraints simultaneously for a model with universal soft-supersymmetry breaking parameters.

We compute the B-parameter at the GUT scale here for two models: a generic one-parameter model~\cite{Lopez:1993rm,Lopez:1994fz,Lopez:1995hg,Maxin:2008kp} and an M-Theory model~\cite{TJLI}. We find that for the models with a predicted B-parameter at the GUT scale, namely the minimal one-parameter model and the M-Theory model without corrections, i.e. $x = 0$, contrary to the solutions discovered when only considering the experimental constraints, there are no solutions when the B-parameter constraint is taken into account. In light of this, we shall vary the moduli for the one-parameter model to investigate whether some solutions can be found that satisfy the B-parameter constraint, in addition to only satisfying the experimental constraints. It is also necessary to determine whether solutions exist for the M-Theory model when $x\neq0$ that can satisfy the B-parameter constraint. 

For the one-parameter model, we begin with the minimal model in the special dilaton scenario with the soft terms of the form

\begin{equation}
m_{0} = \frac{1}{\sqrt{3}}m_{1/2}, \ \ \ \ \ A = -m_{1/2}, \ \ \ \ \ B = -\frac{2}{\sqrt{3}}m_{1/2}.
\end{equation}

\noindent and construct a method of varying the modular weights. To accomplish this, we modify the expressions above and introduce three new parameters $\xi, \delta,$ and $\eta$ that will allow us to investigate more general cases:

\begin{equation}
\label{eqn:dilaton_nm}
m_{0} = \frac{\xi}{\sqrt{3}}m_{1/2}, \ \ \ \ \ A = -\delta m_{1/2}, \ \ \ \ \ B = -\frac{\eta}{\sqrt{3}}m_{1/2}.
\end{equation}

\begin{figure}[htp]
	\centering
		\includegraphics[width=0.49\textwidth]{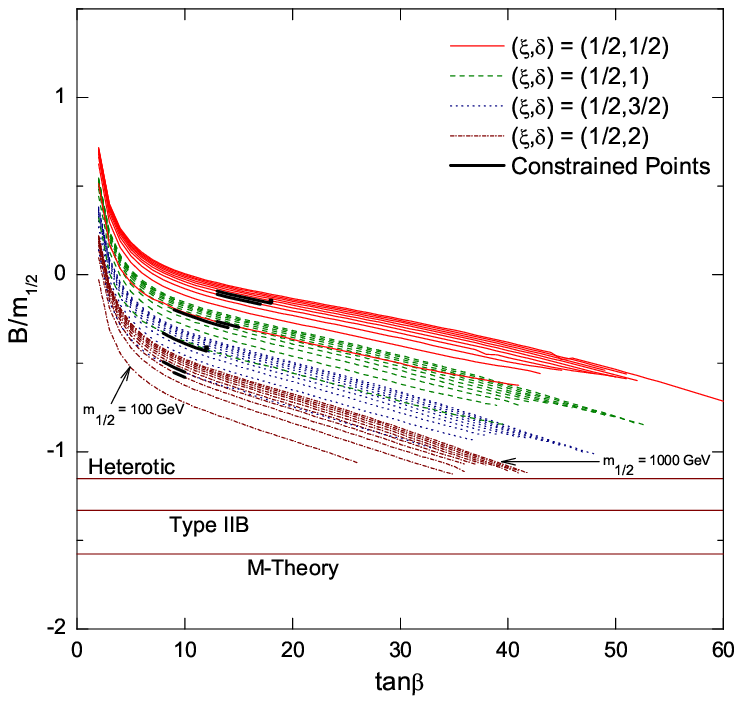}
		\includegraphics[width=0.49\textwidth]{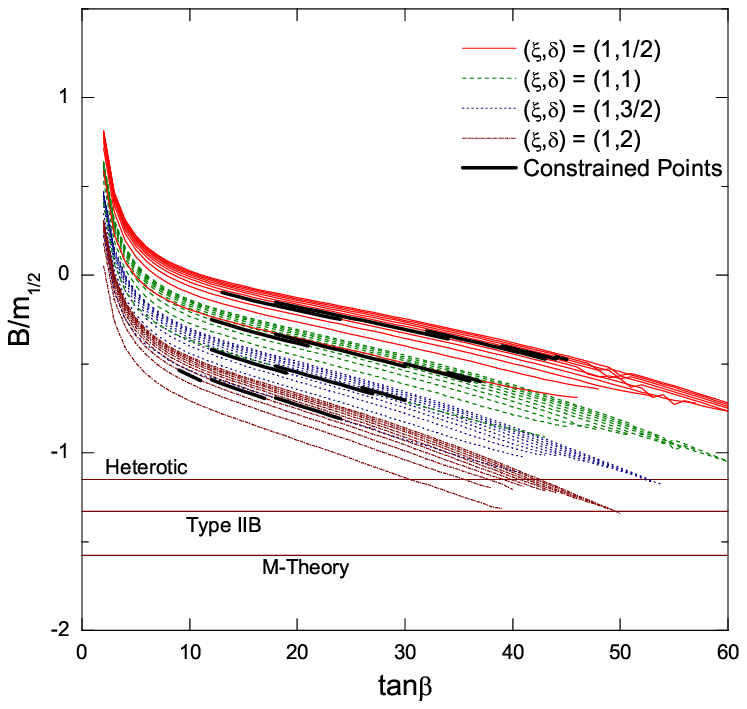}
		\includegraphics[width=0.49\textwidth]{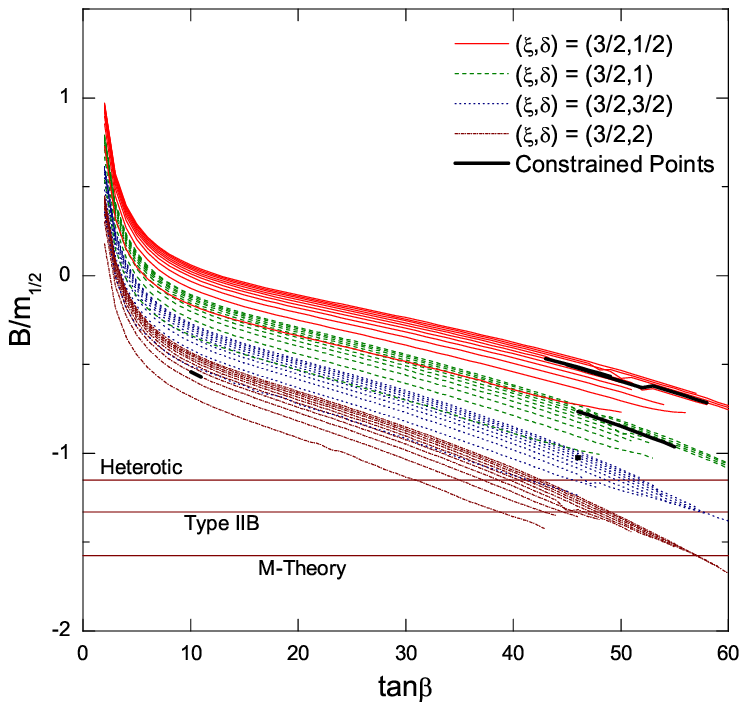}
		\includegraphics[width=0.49\textwidth]{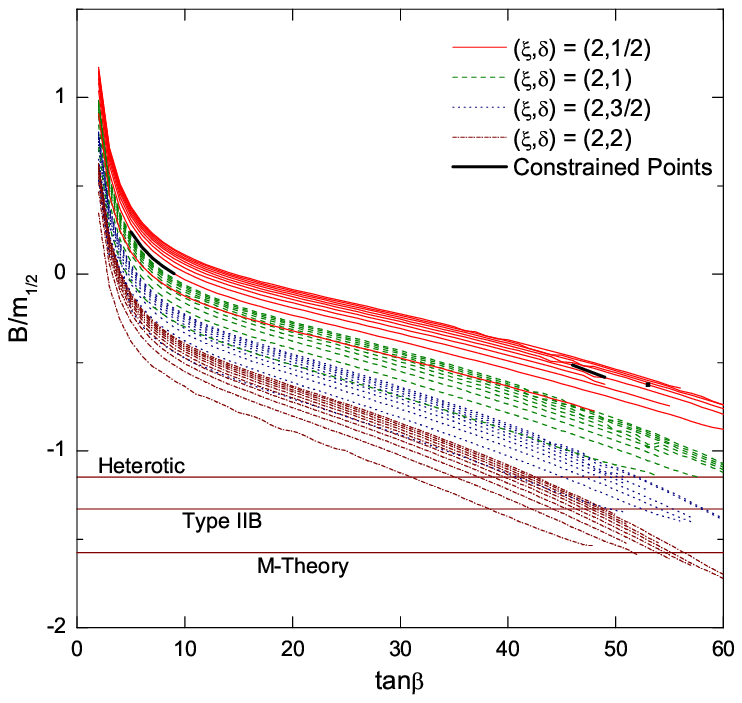}
		\caption{B/$m_{1/2}$ vs. tan$\beta$ at the GUT scale for the one-parameter model. Each plot contains four sets of ten curves, each set with a different $(\xi,\delta)$. The ten curves are for $m_{1/2}$ = 100, 200,..., 900, 1000 GeV, where the lowermost curve in each set is $m_{1/2}$ = 100 GeV and the uppermost curve in each set is $m_{1/2}$ = 1000 GeV. The three horizontal lines represent the predictions for B at the GUT scale. The segments of the curves highlighted in thick black represent those points in the parameter space which are experimentally allowed. The minimal one-parameter model and the M-Theory model with $x=0$ is the case $(\xi,\delta)=(1,1)$. In these plots, all the allowed points highlighted in black satisfy the relic neutralino density in the SSC scenario. Those points satisfying only the WMAP relic density are not highlighted. As the plots show, the points experimentally allowed do not intersect the predictions for B, hence, the B-parameter constraint cannot be satisfied by the models displayed in this Figure.}
	\label{fig:B_OPM}
\end{figure}

Using these expressions, the minimal one-parameter model is the case $(\xi,\delta)=(1,1)$. We now let $\xi=\frac{1}{2},1,\frac{3}{2},2$ and $\delta=\frac{1}{2},1,\frac{3}{2},2$, which will give us 16 different cases to examine. The 16 cases shall be divided up into four data sets such that each data set will be plotted independently. Each data set will have constant $\xi$, and thus constant $m_{0}$, while $\delta$ is varied, and hence $A$ is varied. Therefore, each of the four plots will contain four sets of curves, where each set of curves pertains to one $(\xi,\delta)$. The gaugino mass is incremented from 100 GeV to 1000 GeV in steps of 100 GeV, whereas tan$\beta$ is varied in increments of one from 2 to 60. From these specifications, a list of soft supersymmetry breaking terms is generated and the B-parameter at the GUT scale is calculated for each set of soft terms. As shown in Fig.~\ref{fig:B_OPM}, there are solutions to the one-parameter model when only the experimental constraints are considered, though when the B-parameter constraint is applied, the experimentally allowed parameter space is nullified. There are no intersections between the B-parameter curves and the horizontal lines representing the predictions for the B-parameter. Note that $\eta$ for the three predictions are

\begin{description}
 \centering
  \item[ ] $\eta = 2 ~~(heterotic)$
  \item[ ] $\eta = \frac{4}{\sqrt{3}} ~~(Type IIB)$
  \item[ ] $\eta = 1 + \sqrt{3} ~~(M-Theory)$
\end{description}

To further ensure that we have examined all possibilities for the minimal one-parameter model, we computed an additional case with an independent modular weight for the Higgs scalars at the unification scale. Our motivation for attempting this is that since the Higgs typically come from a different sector, the dependence on the Kahler moduli should be different. While keeping $(\xi,\delta)=(1,1)$, the modular weight on the Higgs scalar was varied, nonetheless, there was no shifting of the B-parameter curves and only a slight change in the number of points allowed by the experimental constraints. Lastly, we varied the stop mass at the unification scale for the minimal one-parameter model case $(\xi,\delta)=(1,1)$ in an attempt to find solutions allowed by the experimental constraints that can also meet the B-parameter constraint, however there were no solutions in this case either. Therefore, for the minimal one-parameter model parameterizations, the B-parameter constraint at the scale $M_{High} = M_{GUT}$ cannot be satisfied.

We now look more closely at an M-Theory model by varying the unknown parameter $x$, and due to restrictions on the gauge coupling, seek solutions only for $0 \leq x \leq 1$. The angle $\theta$ in the expressions in~\cite{TJLI} can also represent an unknown parameter, but we choose to let it remain constant for our study here and only vary the parameter $x$. The M-Theory expressions are given in terms of the gravitino mass $m_{3/2}$, so first the relations for the soft terms must be solved in terms of $m_{1/2}$, and these are

\newpage
\begin{description}
 \centering
  \item[ ] $m_{0} = (1+x) \sqrt{\frac{1}{3} - \frac{x(6+x)}{(3+x)^{2}}}m_{1/2}$
  \item[ ] $A = -\frac{(3-2x)(1+x)}{(3+x)}m_{1/2}$
  \item[ ] $B = -\frac{[3 + 3\sqrt{3}-(\sqrt{3}-1)x](1+x)}{\sqrt{3}(3+x)}m_{1/2}$
\end{description}

\begin{figure}[htp]
	\centering
		\includegraphics[width=0.49\textwidth]{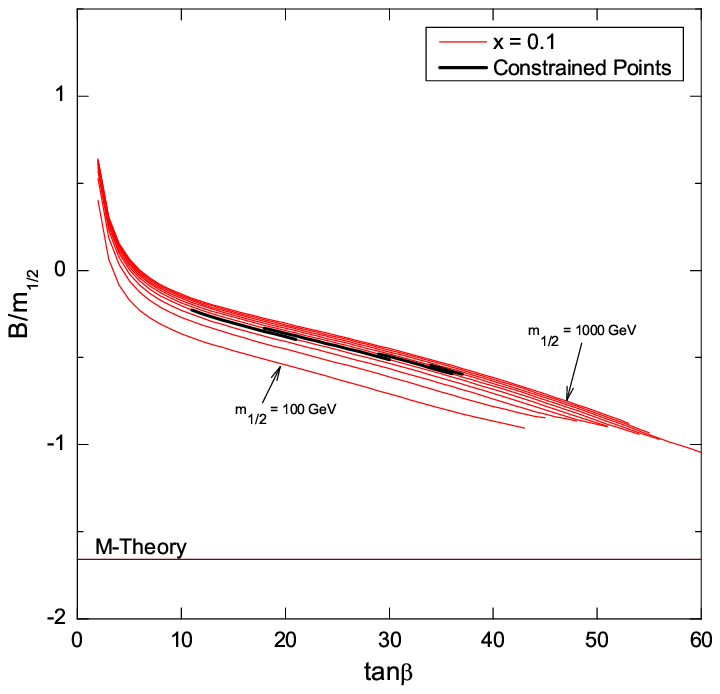}
		\includegraphics[width=0.49\textwidth]{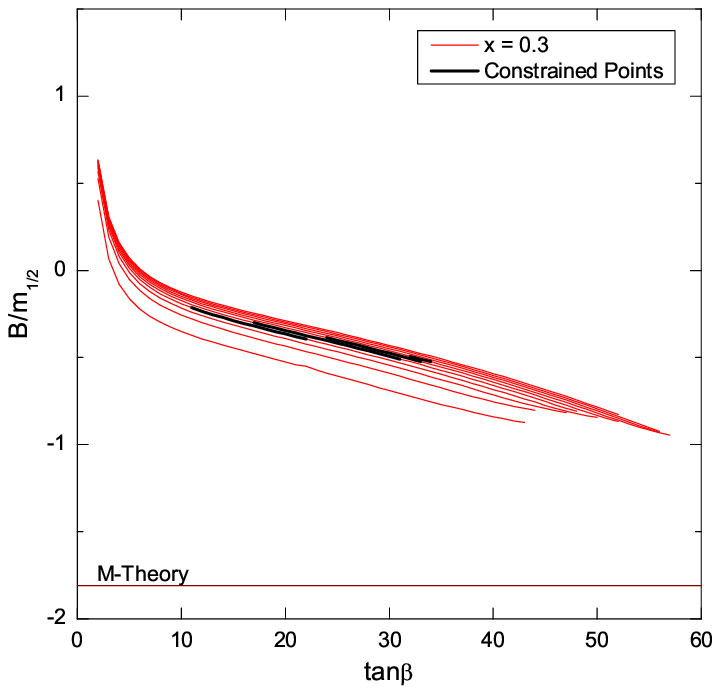}
		\includegraphics[width=0.49\textwidth]{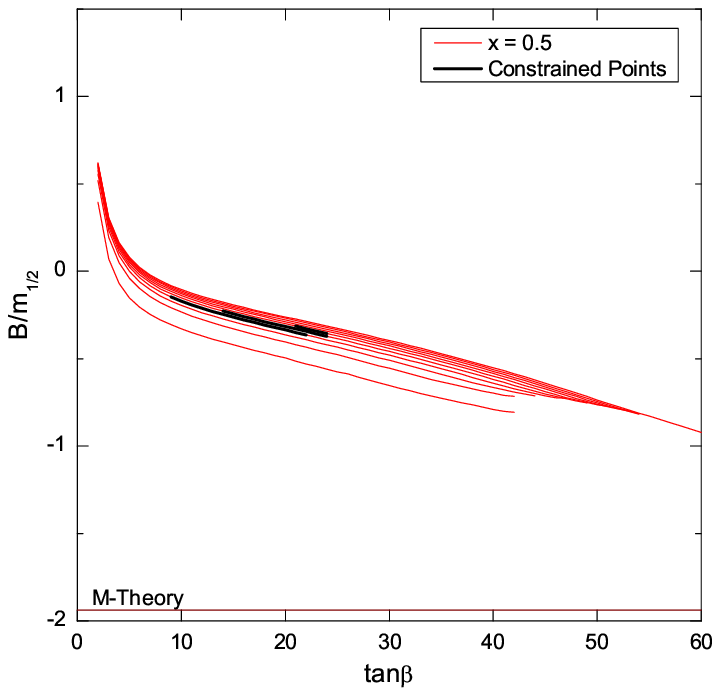}
		\caption{B/$m_{1/2}$ vs. tan$\beta$ at the GUT scale for an M-Theory model. Each plot contains a set of curves for a different $x$. The ten curves in each plot are for $m_{1/2}$ = 100, 200,..., 900, 1000 GeV, where the lowermost curve in each plot is $m_{1/2}$ = 100 GeV and the uppermost curve in each plot is $m_{1/2}$ = 1000 GeV. The horizontal lines represent the predictions for B at the GUT scale. The segments of the curves highlighted in thick black represent those points in the parameter space which are experimentally allowed. In these plots, all the allowed points highlighted in black satisfy the relic neutralino density in the SSC scenario. Those points satisfying only the WMAP relic density are not highlighted. As the plots show, the points experimentally allowed do not intersect the predictions for B, hence, the B-parameter constraint cannot be satisfied by the model displayed in this Figure.}
	\label{fig:B_MTheory}
\end{figure}

\vspace{1cm}
We scan for real solutions that give $m_{0} > 0$, $A < 0$, and $B < 0$, and find these solutions only exist for $0 \leq x \leq 0.6742$. The case $x=0$ is shown in Fig.~\ref{fig:B_OPM}, so we run the three additional cases $x=0.1,0.3,0.5$ for the same increments of $m_{1/2}$ and tan$\beta$ as the one-parameter model, and compute the allowed parameter space from the experimental constraints. Again, even though there are points allowed within the parameter space when only considering the experimental constraints, none of these allowed points can satisfy the B-parameter constraint for $0 \leq x \leq 0.6742$. This is clearly shown in Fig.~\ref{fig:B_MTheory} where the horizontal line representing the prediction for the B-parameter at the GUT scale does not intersect the B-parameter curves in any of the three sample cases. In fact, as the unknown variable $x$ increases toward the upper end of its range, the discrepancy becomes larger. Here again, as in the one-parameter model, the M-Theory model cannot produce any viably allowed parameter space that satisfies both the experimental constraints and the B-parameter constraint.

\section{Conclusion}
A well-motivated framework for studying supersymmetry breaking is to assume that it is dominated by the K\a"ahler moduli and/or the
universal dilaton.  Such scenarios give rise to very constrained supersymmetry breaking soft-terms which depend only on a universal gaugino mass. In addition, modulus-dominated supersymmetry breaking appears as a generic feature of many string compactifications.  
We find that the simplest models are not viable, at least under a standard RGE running between the electroweak scale 
and $M_{GUT}=2\times 10^{16}$~GeV. Although these models may have some parameter space which can satisfy experimental constraints, the value of tan$\beta$ determined at the electroweak scale is not consistent with the $B$ parameter at the GUT scale.  Despite this, it is still possible that supersymmetry breaking could be dominated by
the moduli if one considers a non-standard RGE running or if the high-energy scale $M_{High}$ at which the boundary condition on the
soft-terms is defined is different from $M_{GUT}$.  A non-standard RGE running could result if vector-like matter is introduced at
intermediate mass scales.  Indeed, the introduction of such vector-like matter is one way of pushing the GUT scale up to the string scale
$M_{string}=\mathcal{O}(10^{18})$~GeV.  Moreover, the string scale for large-volume Type IIB flux models can be substantially lower than $M_{GUT}$. Thus, 
modulus-dominated supersymmetry breaking is possibly still viable under non-minimal assumptions.  It would be very interesting to study
such scenarios and we plan to return to this question in future work.   

\section{Acknowledgments}

This research was supported in part by the Mitchell-Heep Chair in High Energy Physics and by DOE grant DE-FG03-95-Er-40917. The work of
VEM is supported by the U.S. National Science Foundation under grant PHY-0757394.

\end{document}